\documentstyle[preprint,aps,epsfig]{revtex}

\input{epsf}
\input{psfig}
\draft
\begin{document}
\title{Dynamic Structure Factor of Liquid
and Amorphous Ge From Ab Initio Simulations}
\author{Jeng-Da Chai\footnote{Present address:
Institute for Physical Science and Technology, University of Maryland,
College Park, Maryland 20742; email jdchai@wam.umd.edu}
and D. Stroud\footnote{Corresponding author: tel.
(614)292-8140; fax (614)292-7557; email stroud@mps.ohio-state.edu}}
\address{Department of Physics,
The Ohio State University, Columbus, Ohio 43210}
\author{J. Hafner and G. Kresse}
\address{Institut f\"{u}r Materialphysik and Center for
Computational Materials Science, Technische Universit\"{a}t
Wien, Sensengasse 8/12, A-1090 Wien, Austria}

\date{\today}

\maketitle

\begin{abstract}

We calculate the dynamic structure factor 
$S(k, \omega)$ of liquid Ge ($\ell$-Ge) at temperature $T = 1250$ K,
and of amorphous Ge ($a$-Ge) at $T = 300$ K, using 
{\em ab initio} molecular dynamics.
The electronic energy is computed using density-functional
theory, primarily in the generalized gradient approximation,
together with a
plane wave representation of the wave functions and ultra-soft
pseudopotentials.  We use a 64-atom cell with periodic
boundary conditions, and calculate averages over runs of
up to about 16 ps.    
The calculated liquid $S(k, \omega)$ agrees qualitatively
with that obtained by Hosokawa {\it et al}, using inelastic
X-ray scattering.  In $a$-Ge, we find that
the calculated $S(k, \omega)$ is in qualitative agreement with
that obtained experimentally by Maley {\it et al}.
Our results suggest that the {\em ab initio} approach is 
sufficient to allow approximate calculations of $S(k, \omega)$ in 
both liquid and amorphous materials.   
\end{abstract}



\section{INTRODUCTION}

Ge is a well-known semiconductor in its solid phase, but becomes
metallic in its liquid phase.  Liquid Ge ($\ell$-Ge) has, near
its melting point, an electrical conductivity characteristic of
a reasonably good metal 
($\sim 1.6 \times 10^{-4}\Omega^{-1}$cm$^{-1}$\cite{glazov}), but it retains
some residual structural features of the solid semiconductor.
For example, the static structure factor $S(k)$, has a shoulder
on the high-$k$ side of its first (principal) peak, which is
believed to be due to a residual tetrahedral short-range order.
This shoulder is absent in more conventional liquid metals such
as Na or Al, which have a more close-packed structure in
the liquid state and a shoulderless first peak in the structure
factor.  Similarly, the bond-angle distribution function just
above melting is believed to have peaks at {\em two} angles,
one near $60^o$ and characteristic of close packing, and one
near $108^o$, indicative of tetrahedral short range order.
This latter peak rapidly disappears with increasing temperature
in the liquid state.

These striking properties of $\ell$-Ge have been studied
theoretically by several groups.
Their methods fall into two broad classes: empirical and
first-principles.  A typical empirical calculation is that
of Yu {\it et al}\cite{yu}, who calculate the structural properties
of $\ell$-Ge assuming that the interatomic potentials in $\ell$-Ge
are a sum of two-body and three-body potentials of the form
proposed by Stillinger and Weber\cite{sw}.  These authors
find, in agreement with experiment, that there is a high-k
shoulder on the first peak of S(k) just above melting, which
fades away with increasing temperature.  However, since in this
model {\em all} the potential energy is described by a sum
of two-body and three-body interactions, the interatomic
forces are probably stronger, and the ionic
diffusion coefficient is correspondingly smaller, than their
actual values.

In the second approach, the electronic degrees of freedom are
taken explicitly into account.  If the electron-ion interaction
is sufficiently weak, it can be treated by linear response
theory\cite{as}.  In linear response, the total energy in
a given ionic configuration is a term which is independent of
the ionic arrangement, plus a sum of two-body ion-ion effective
interactions.  These interactions typically do not give the
bond-angle-dependent forces which are present in the experiments,
unless the calculations are carried to third order in the
electron-ion pseudopotential\cite{as}.
Such interactions are, however, included in the so-called
{\em ab initio} approach, in which the forces on the ions are
calculated from first principles, using the
Hellman-Feynman theorem together with density-functional
theory\cite{kohn1} to treat the energy of the inhomogeneous electron gas.  This
approach not
only correctly gives the bond-angle-dependent ion-ion interactions,
but also, when combined with standard molecular dynamics
techniques, provides a good account of the electronic properties
and such dynamical ionic properties as the ionic self-diffusion
coefficients.

This combined approach, usually known as {\em
ab initio} molecular dynamics, was pioneered by Car and
Parrinello\cite{car}, and, in somewhat different form, has
been applied to a wide range of liquid metals and alloys,
including $\ell$-Ge\cite{kresse,takeuchi,kulkarni},
$\ell$-Ga$_x$Ge$_{1-x}$\cite{kulkarni1}, stoichiometric
III-V materials such as $\ell$-GaAs, $\ell$-GaP, and
$\ell$-InP\cite{zhang,lewis}, nonstoichiometric
$\ell$-Ga$_x$As$_{1-x}$\cite{kulkarni2},
$\ell$-CdTe\cite{cdte}, and $\ell$-ZnTe\cite{znte}, among
other materials which are semiconducting in their solid
phases.  It has been employed to calculate a wide
range of properties of these materials, including the static
structure factor, bond-angle distribution function, single-particle
electronic density of states, d.\ c.\ and a.\ c.\ electrical
conductivity, and ionic self-diffusion coefficient.
The calculations generally agree quite well with
available experiment.

A similar {\em ab initio} approach has also been applied extensively
to a variety of amorphous semiconductors, usually obtained by
quenching an equilibrated liquid state from the melt.
For example, Car, Parrinello and their collaborators
have used their own {\em ab initio} approach (based on
treating the Fourier components of the electronic wave functions
as fictitious classical variables) to obtain many structural
and electronic properties of amorphous Si\cite{car1,stich}.
A similar approach has been used by Lee and Chang\cite{lee}.
Kresse and Hafner\cite{kresse} obtained both S(k) and g(r),
as well as many electronic properties, of $a$-Ge, using an
{\em ab initio} approach similar to that used here, in which
the forces are obtained directly from the Hellmann-Feynman theorem
and no use is made of fictitious dynamical variables for the
electrons, as in the Car-Parrinello approach.  A similar calculation
for $a$-Si has been carried out by Cooper {\it et al}\cite{cooper},
also making use of a plane wave basis and treating the electron
density functional in the generalized gradient approximation
(GGA)\cite{gga}.  More recently, a number of calculations for $a$-Si
and other amorphous semiconductors have been carried out by
Sankey {\it et al}\cite{sankey}, and by Drabold and collaborators
\cite{drabold}.  These calculations use {\em ab initio} molecular
dynamics and electronic density functional theory, but in a
localized basis.  A recent study, in which S(k) and g(r) were
computed for several {\em ab initio} structural models of
$a$-Si, has been carried out by Alvarez {\it et al}\cite{alvarez}.

Finally, we mention a third approach, intermediate between 
empirical and {\em ab initio} molecular dynamics, and generally
known as tight-binding molecular dynamics.  In this approach, 
the electronic part of the total energy is described using
a general tight-binding Hamiltonian for the band electrons.  The
hopping matrix elements depend on separation between the ions,
and additional terms are included to account for the various
Coulomb energies in a consistent way.  The parameters can be
fitted to {\em ab initio} calculations, and forces on the
ions can be derived from the separation-dependence of the hopping
matrix elements.  This approach has been used, e. g., to treat
$\ell$-Si\cite{wang1}, $a$-Si\cite{servalli}, and 
liquid compound semiconductors such as $\ell$-GaAs and 
$\ell$-GaSb\cite{molteni}.  Results
are in quite good agreement with experiment.

In this paper, we extend the method of {\em ab initio} molecular
dynamics to another dynamical property of the ions: the {\em
dynamical structure factor}, denoted $S(k, \omega)$.  While
no fundamentally new theory is required to calculate
$S(k, \omega)$, this quantity provides additional information about
the time-dependent ionic response beyond what can be extracted
from other quantities.  The present work appears to be the
first to calculate $S(k, \omega)$ using {\em ab initio}
molecular dynamics.  Here, we will calculate $S(k,\omega)$
for $\ell$-Ge, where some recent experiments\cite{hosokawa}
provide data for comparison, and also for amorphous
Ge ($a$-Ge).  In the latter case, using a series of
approximations described below,
we are able to infer the {\em vibrational} density of states
of as-quenched $a$-Ge near temperature $T = 300$K in reasonable
agreement with experiment.  The calculated
$S(k, \omega)$ for the liquid also
agrees quite well with experiment,
especially considering the computational uncertainties inherent in
an {\em ab initio} simulation with its necessarily small number
of atoms and limited time intervals.

The remainder of this paper is organized as follows.  A brief review
of the calculational method is given in Section II.  The results
are presented in Section III, followed by a discussion and
a summary of our conclusions in Section IV.

\section{METHOD}

Our method is similar to that described in several
previous papers\cite{kulkarni,kulkarni1,kulkarni2}, but uses the 
Vienna Ab Initio Simulation Package (VASP), whose workings have been extensively
described in the literature\cite{kresse1}.  Briefly, the calculation
involves two parts.  First, for a given ionic
configuration, the total electronic energy is
calculated, using an approximate form of the
Kohn-Sham free energy density functional, and the
force on each ion is also calculated, using the
Hellmann-Feynman theorem.  Second, Newton's equations
of motion are integrated numerically for the ions,
using a suitable time step.  The process is repeated
for as many time steps as are needed to calculate
the desired quantity.  To hold the temperature constant,
we use the canonical ensemble with the velocity
rescaled at each time step.  Further details of this
approach are given in Ref. \cite{kulkarni}.

The VASP code uses ultrasoft Vanderbilt 
pseudopotentials\cite{vanderbilt}, a plane wave basis for
the wave functions, with the original
Monkhorst-Pack ($3 \times 3 \times 3$)
k-space meshes\cite{monkhorst} and a total of 21,952 plane waves, corresponding
to an energy cut-off of 104.4eV. 
We use a finite-temperature
version of the Kohn-Sham theory\cite{mermin} in which the
electron gas Helmholtz free energy is calculated on each time
step.  This version also
broadens the one-electron energy levels to make the
k-space sums converge more rapidly.  Most of our
calculations are done using the generalized gradient
approximation (GGA)\cite{gga} for the exchange-correlation
energy (we use the particular form of the GGA developed by 
Perdew and Wang\cite{gga}), 
but some are also carried out using
the local-density approximation (LDA).

In our iteration of Newton's Laws in liquid Ge
($\ell$-Ge), we typically start from the diamond
structure (at
the experimental liquid state density for the
temperature of interest), then iterate
for 901 time steps, each of 10 fs, using the LDA.  To obtain
S(k) within the GGA, we start from the LDA configuration
after 601 time steps, then iterate using the GGA for an
additional 1641 10-fs time steps, or 16.41 ps.
We calculate the GGA S(k) by averaging over an interval
of 13.41 ps within this 16.41 time interval, starting a
time $t_2$ after the start of the GGA simulation.  We average
over all $t_2$'s from 1.0 to 3.0 ps.

For comparison, we have also calculated S(k) within the LDA.
This S(k) is obtained by averaging over 601 time steps of
the 901 time-step LDA simulation.  This 601-step interval
is chosen to start a time $t_1$ after the start of this
simulation; the calculated LDA S(k) is also averaged over
all $t_1$'s from 1ps to 3ps.    

To calculate quantities for amorphous Ge (a-Ge), we
start with Ge in the diamond structure at T =
1600K and at the experimental density for that
temperature, as quoted in Ref. \cite{kulkarni}.  
Next, we quench this
sample to 300 K, cooling at a uniform rate over,
so as to reach 300 K in about 3.25 ps (in 10-fs time steps).  
Finally, starting from T = 300 K, 
we iterate
for a further 897 time steps, each of 10 fs, or 8.97 ps, 
using the LDA.
The LDA S(k) is then obtained by averaging over 5.97 of those
8.97 ps, starting a time $t_1$ after the system is reached
300K; we also average this S(k) over all $t_1$'s from 1.0 to
3.0 ps.  To obtain S(k) within the GGA, we start the GGA after
5.7 ps of the LDA simulation, 
and then iterate using the GGA for an additional 18.11
ps in 10-fs time steps.  The GGA S(k) is obtained by averaging
over a 15.11 ps time interval of this 18.11 ps run, starting
a time $t_2$ after the start of the GGA simulation; we also
average over all values of $t_2$ from 1.0 to 3.0 ps. 

The reader may be concerned that the 3.25 ps quench time is very
short, very unrepresentative of a realistic quench, 
and very likely to produce numerous defects in the quenched 
structure,
which are not typical of the $a$-Ge studied in experiments.
In defense, we note that some of these defects may be annealed
out in the subsequent relaxation at low temperatures, which is
carried out before the averages are taken.  In addition, the
static structure factor we obtain agrees well with
experiment, and the dynamic structure factor is also
consistent with experiment, as discussed below.  Thus, the
quench procedure appears to produce a material rather similar
to some of those studied experimentally.  Finally, we note that
experiments on $a$-Ge themselves show some variation, 
depending on the exact method of sample preparation.

We have used the procedure outlined above to calculate various
properties of $\ell$-Ge and $a$-Ge.  Most of
these calculated properties have been described
in previous papers, using slightly
different methods, and therefore will be discussed here only
very briefly\cite{compare}.  However, our results for the dynamic
structure factor $S(k, \omega)$ as a function of
wave vector k and frequency $\omega$ are new, and will
be described in detail.  We also present
our calculated static structure factor $S(k)$, which
is needed in order to understand the dynamical 
results.

$S(k)$ is defined by the relation
\begin{equation}
S(k) = \frac{1}{N}\langle \sum_{i,j}\exp[i{\bf k}\cdot({\bf r}_i(t)
-{\bf r}_j(t))]\rangle_{t_0} - N\delta_{{\bf k}, 0},
\label{eq:static}
\end{equation}
where ${\bf r}_i$ is the position of the i$^{th}$ ion at time
t, $N$ is the number of ions in the sample, and the
triangular brackets denote an average over the
sampling time.  In all our calculations, we have
used a cubic cell with $N = 64$ and periodic boundary
conditions in all three directions.  This choice of
particle number and cell shape is compatible with
any possible diamond-structure Ge within the 
computational cell.

$S({\bf k}, \omega)$ is defined by the relation (for 
${\bf k} \neq 0$, $\omega \neq 0$)
\begin{equation}
S({\bf k}, \omega) = \frac{1}{2\pi N}\int_{-\infty}^\infty
\exp(i\omega t)\langle \rho({\bf k},t)\rho(-{\bf k},0)\rangle dt,
\label{eq:defskw}
\end{equation}
where the Fourier component $\rho({\bf k}, t)$ of the number
density is defined by
\begin{equation}
\rho({\bf k}, t) = \sum_{i = 1}^N\exp(-i{\bf k}\cdot{\bf r}_i(t)).
\label{eq:rhokt}
\end{equation}
In our calculations, the average $\langle...\rangle$ is computed as
\begin{equation}
\langle \rho({\bf k}, t)\rho({\bf -k},0)\rangle
= \frac{1}{\Delta t_1}\int_0^{\Delta t_1}
\rho({\bf k}, t_1 + t)\rho({\bf -k},t_1)dt_1
\end{equation}
over a suitable range of initial times $t_1$.  Because of the
expected isotropy of the liquid or amorphous phase, which should
hold in the limit of large N, $S({\bf k}, \omega)$ should 
be a function only of the magnitude $k$ rather than the
vector ${\bf k}$, as should the structure factor $S({\bf k})$.

Our calculations are carried out over relatively
short times.  To reduce statistical errors, we 
therefore first calculate
\begin{equation}
S({\bf k}, \omega, t_1, t_2) = \frac{1}{\pi N}\int_{t_1}^{t_2}
dt \rho({\bf k},t_1 + t)\rho(-{\bf k}, t_1)\exp(i\omega t).
\end{equation}
For large enough $t_2$, $S({\bf k}, \omega, t_1, t_2)$
should become independent of $t_2$ but will still
retain some dependence on $t_1$.   Therefore, in the liquid, we
obtain our calculated dynamic structure factor,
$S_{calc}({\bf k}, \omega)$, by averaging over a
suitable range of $t_1$ from 0 to $\Delta t_1$:
\begin{equation}
S_{calc}({\bf k},\omega) = \frac{1}{\Delta t_1}
\int_{0}^{\Delta t_1}dt_1 S({\bf k},
\omega, t_1, t_2).
\end{equation}
We choose our initial time in the $t_1$ integral to be 1 ps
after the start of the GGA calculation, and (in the liquid)
$\Delta t_1 = 7$ ps.   We choose the final MD time $t_2 = 16.41$ps. 
For $a$-Ge, we use the same procedure but $t_2 = 18.11$ps 
in our simulations.
For our finite simulational sample, the calculated
$S({\bf k})$ and $S({\bf k}, \omega)$ will, in fact,
depend on the direction as well as the magnitude of
${\bf k}$.  To suppress this finite-size effect, we
average the calculated $S({\bf k})$ and $S({\bf k}, \omega)$ 
over all ${\bf k}$ values of the same
length.  This averaging considerably reduces 
statistical error in both $S(k,\omega)$ and $S(k)$.

Finally, we have also incorporated the experimental
resolution functions into our plotted values of
$S(k,\omega)$.  Specifically, we generally plot
$S_{av}(k,\omega)/S(k)$, where $S_{av}(k,\omega)$ is
obtained from the (already orientationally averaged)
$S(k,\omega)$ using the formula
\begin{equation}
S_{av}(k, \omega) = \int_{-\infty}^\infty R(\omega - \omega^\prime)
S(k, \omega^\prime)d\omega^\prime,
\label{eq:resol}
\end{equation}
where the resolution function $R(\omega)$ (normalized so
that $\int_{-\infty}^\infty R(\omega)d\omega = 1$) is
\begin{equation}
R(\omega) = \frac{1}{\sqrt{\pi}\omega_0}\exp(-\omega^2/\omega_0^2).
\label{eq:resol1}
\end{equation}

In an isotropic liquid, we must have 
$S(k, -\omega) = S(k, \omega)$, since our ions are assumed to
move classically under the calculated first-principles force.   
Our orientational
averaging procedure guarantees that this will be
satisfied identically in our calculations, since
for every ${\bf k}$, we always include ${\bf -k}$ in 
the same average.  We will nonetheless show results
for negative $\omega$ for clarity, but they do not
provide any additional information.

\section{RESULTS}

\subsection{S(k) for $\ell$-Ge and $a$-Ge}

In Fig.\ 1, we show the calculated $S(k)$ for
$\ell$-Ge at $T = 1250$ K, as obtained using the procedure
described in Section II.  The two calculated
curves are obtained using the GGA and the LDA
for the electronic energy-density functional; they
lead to nearly identical results.  The calculated
$S(k)$ shows the well-known characteristics already
found in previous simulations\cite{kresse,kulkarni}.  Most notably,
there is a shoulder on the high-$k$ side of the
principal
peak, which is believed to arise from residual
short-range tetrahedral order persisting into the
liquid phase just above melting.  We also show the
experimental results of Waseda {\it et al}\cite{waseda}; agreement
between simulation and experiment is good, and in 
particular the shoulder seen in experiment is also
present in both calculated curves (as observed also
in previous simulations).

We have also calculated
S(k) for a model of {\em amorphous} Ge ($a$-Ge) at $300 K$.  Our
sample of $a$-Ge is prepared as follows.  First, we create an
equilibrated sample of $\ell$-Ge at a temperature of $1600 K$
and the experimental density corresponding to that
temperature (as quoted in Ref.\ \cite{kulkarni}).
Next, we quench this sample of $\ell$-Ge to $300 K$, by
cooling at a uniform rate over a time interval
of about 3.25ps.  Finally, starting
from $T = 300 K$, we evaluate the structure factor by averaging
over 800 time steps of 10 fs each, or 8 ps, after first
discarding 100 time steps at $T = 300$K.   We also average
S(k) over different ${\bf k}$ vectors of
the same length, as for $\ell$-Ge.
We use the measured number density of $0.04370 \AA^{-3}$ 
for $a$-Ge at $T = 300 K$\cite{etherington}.

In Fig.\ 2, we show the calculated $S(k)$ for 
$a$-Ge at T = 300, again using both the GGA and the LDA.
The sample is prepared and the averages obtained as
described in Section II.  In contrast to $\ell$-Ge,
but consistent with previous simulations\cite{kresse,alvarez},
the principal peak in S(k) is strikingly split.
The calculations are in excellent agreement with
experiments carried out on as-quenched $a$-Ge 
at $T = 300$ K\cite{etherington}; in particular, the split principal
peak seen in experiment is accurately reproduced by
the simulations.

We have also calculated a number of other quantities for
both $\ell$-Ge and $a$-Ge, including pair distribution
function $g(r)$, and the electronic density of
states $n(E)$.  

For $\ell$-Ge, we calculated $n(E)$ using the Monkhorst-Pack
mesh with gamma point shifting (one of the meshes recommended
in the VASP package).  The resulting $n(E)$
is generally similar to that found in previous 
calculations\cite{kresse,kulkarni}, provided that an average is
taken over at least 5-10 liquid state configurations.
Our $n(E)$ for $a$-Ge [calculated
using a shorter averaging time than that used below for
$S(k, \omega)$] is also similar to that found previously\cite{kresse}.
Our calculated $g(r)$'s for both $\ell$-Ge and $a$-Ge, 
as given by the VASP program,
are similar to those found in Refs.\ \cite{kresse} and 
\cite{kulkarni}.  The calculated 
number of nearest neighbors in the first
shell is 4.18 for $a$-Ge 
measured to the first minimum after the principal peak in g(r).
For $\ell$-Ge, if we count as ``nearest neighbors''
all those atoms within 3.4\AA of the central atom
(one of the cutoffs used in Ref.\ \cite{kresse})
we find approximately 6.1 nearest neighbors, very close to
the value of 5.9 obtained in Ref.\ \cite{kresse} for that
cutoff.  
Finally, we have recalculated the self-diffusion coefficient
$D(T)$ for $\ell$-Ge at $T = 1250$ K, from the time derivative
of the calculated mean-square ionic displacement; we 
obtain a result very close to that of Ref.\ \cite{kulkarni}.

\subsection{$S(k,\omega)$ for $\ell$-Ge and $a$-Ge}

\subsubsection{$\ell$-Ge}

Fig.\ 3 shows the calculated ratio $S(k,\omega)/S(k)$ for
$\ell$-Ge at $T = 1250$ K, as obtained using the
averaging procedure described in Section II.  We
include a resolution function [eqs. (\ref{eq:resol}) and 
(\ref{eq:resol1})]
of width $\hbar\omega = 2.5$ meV, the same as the
quoted experimental width\cite{hosokawa}.  In Fig.\ 4, we show
the same ratio, but without the resolution function
(i. e., with $\omega_0 = 0$).  Obviously, there is
much more statistical noise in this latter case,
though the overall features can still be 
distinguished.     

To interpret these results, we first compare the calculated
$S(k, \omega)$ in $\ell$-Ge with hydrodynamic predictions,
which should be appropriate at small $k$ and $\omega$.  The
This prediction
takes the form (see, for example, Ref.\ \cite{hansen}):
\begin{eqnarray}
2\pi\frac{S(k,\omega)}{S(k)} & = &
\frac{\gamma-1}{\gamma}\left(\frac{2D_Tk^2}{\omega^2 + 
(D_Tk^2)^2}\right) + \nonumber \\
& + & \frac{1}{\gamma}\left(\frac{\Gamma k^2}{(\omega + c_s k)^2 +
(\Gamma k^2)^2} + \frac{\Gamma k^2}
{(\omega -c_sk)^2 + (\Gamma k^2)^2}\right).
\label{eq:hydro}
\end{eqnarray}
Here $\gamma = c_P/c_V$ is the ratio of specific
heats and constant pressure and constant volume,
$D_T$ is the thermal diffusivity, $c_s$ is the adiabatic
sound velocity, and $\Gamma$ is the sound
attenuation constant.  $D_T$ and $\Gamma$ can in turn
be expressed in terms of other quantities.  For
example, $D_T = \kappa_T/(\rho c_P)$, where $\kappa_T$
is the thermal conductivity and $\rho$ is the atomic
number density.  Similarly, 
$\Gamma = \frac{1}{2}\left[a(\gamma - 1)/\gamma + b\right]$,
where $a = \kappa_T/(\rho c_V)$ and $b$ is the kinematic
longitudinal viscosity  (see, for example, Ref.
\cite{hansen}, pp. 264-66).

Eq.\ (\ref{eq:hydro}) 
indicates that
$S(k, \omega)$ in the hydrodynamic regime should
have two propagating peaks centered at
$\omega = \pm c_s k$, and a diffusive peak centered
at $\omega = 0$ and of width determined by $D_T$.  
The calculated $S(k, \omega)/S(k)$ for the
three smallest values of $k$ in Fig.\ 3, does show
the propagating peaks.  We
estimate peak values of $\hbar\omega \sim 10$ meV for
$k = 5.60$ nm$^{-1}$,
$\hbar\omega \sim 11$ meV for $k = 7.92$ nm$^{-1}$,
and (somewhat less clearly) 
$\hbar\omega \sim 13$ meV for $k = 9.70$ nm$^{-1}$.
The value of $c_s$ estimated
from the lowest $k$ value is $c_s \sim 2.7 \times 10^5$ cm/sec.  
(The largest of these three $k$ values may already be outside
the hydrodynamic, linear-dispersion regime.)

These predictions agree reasonably well with the measured
$S(k, \omega)$ obtained by
Hosokawa {\it et al}\cite{hosokawa}, using inelastic X-ray
scattering.
For example, the measured sound-wave peaks for
$k = \pm 6$ nm$^{-1}$ occur near $10$ meV, while those
$k = \pm 12$ nm$^{-1}$
occur at $\hbar\omega = 17.2$ meV, 
Furthermore, the integrated relative
strength of our calculated sound-wave peaks, compared to that of the
central diffusion peak, decreases between
$k = 7.92 $ nm$^{-1}$ and $12.5$ nm$^{-1}$, consistent
with both eq.\ (\ref{eq:hydro})
and the change in experimental behavior\cite{hosokawa} 
between $k = 6$ nm$^{-1}$ and $12$ nm$^{-1}$.

Because $S(k, \omega)$ in Fig.\ 3 already 
includes a significant Gaussian smoothing function, 
a quantitatively accurate half-width for the central peak, and
hence a reliable predicted value for $D_T$, cannot
be extracted.  A rough estimate
can be made as follows.  For the smallest k value of
5.6 nm$^{-1}$, the full width of the central peak at
half-maximum is around $7.5$ meV.  If the only
broadening were due to this Gaussian smoothing, the
full width would be around $2\hbar\omega_0 = 5.5$ meV.
Thus, a rough estimate of the intrinsic full width
is $\approx \sqrt{7.5^2 - 5.5^2} = 5$ meV
$\approx 2\hbar D_Tk^2$.  This estimate
seems reasonable from the raw data
for $S(k, \omega)$ shown in Fig.\ 4.
Using this estimate, one obtains 
$D_T \approx 1.3 \times 10^{-3}$ cm$^2$/sec.  

The hydrodynamic expression for $S(k,\omega)/S(k)$ was
originally obtained without consideration of the
electronic degrees of freedom.  Since
$\ell$-Ge is a reasonably good metal, one might ask
if the various coefficients appearing eq.\ (\ref{eq:hydro})
should be the full coefficients, or just
the ionic contribution to those coefficients.  For
example, should the value of $D_T$ which determines
the central peak width be obtained from the full
$c_P$, $c_V$, and $\kappa_T$, or from only the ionic
contributions to these quantities?  For $\ell$-Ge,
the question is most relevant for $\kappa_T$, since
the dominant contribution to $c_P$ and $c_V$ should be
the ionic parts, even in a liquid metal\cite{as}. 
However, the principal contribution to 
$\kappa_T$ is expected to be the electronic contribution.

We have made an order-of-magnitude estimate of $D_T$
using the experimental liquid number density and the
value $C_P = (5/3)k_B$ per ion, and obtaining the
electronic contribution to $\kappa_T$ from the
Wiedemann-Franz law\cite{am76} together with 
previously calculated estimates of the electronic
contribution\cite{kulkarni}.  This procedure yields 
$D_T \sim 0.1$ cm$^2$/sec, about two orders of magnitude greater
than that extracted from Fig.\ 3, and well outside
the possible errors in that estimate.  We conclude
that the $D_T$ which should be used in eq.\ (\ref{eq:hydro}) for
$\ell$-Ge (and by inference other liquid metals) is the ionic
contribution only.

In support of this interpretation, we consider what
one expects for $S(k, \omega)$ in a simple metal such
as Na.  In such a metal, ionic motions are quite 
accurately determined by effective pairwise 
screened ion-ion interactions\cite{as}.
Since the ionic motion is determined by such an
interaction, the $S(k, \omega)$ resulting from that
motion should not involve the contribution of the
electron gas to the thermal conductivity.  Although
$\ell$-Ge is not a simple metal, it seems plausible
that its $S(k, \omega)$ should be governed by similar
effects, at least in the hydrodynamic regime.  This
plausibility argument is supported by our numerical
results.

For $k$ beyond around $12$ nm$^{-1}$, the hydrodynamic
model should start to break down, since the dimensionless
parameter $\omega\tau$
(where $\tau$ is the Maxwell viscoelastic relaxation time) becomes
comparable to unity.  At these larger $k$'s, both our calculated
and the measured\cite{hosokawa} curves of $S(k,\omega)/S(k)$
continue to exhibit similarities.
Most notable is the existence of a single, rather narrow peak
for $k$ near the principal peak of $S(k)$, followed
by a reduction in height and broadening of this central peak
as $k$ is further increased.  This narrowing was first predicted
by de Gennes\cite{degennes}.  In our calculations, it shows
up in the plot for $k = 20.9$ nm$^{-1}$, for which
the half width of $S(k,\omega)/S(k)$ is quite narrow, while
at $k = 28.5$ and $30.7$ nm$^{-1}$, the corresponding plots
are somewhat broader and lower.   By comparison,
the measured central peak in $S(k,\omega)/S(k)$ is narrow at
$k = 20$ nm$^{-1}$ and especially at $k = 24$ nm$^{-1}$,
while it is broader and lower at $k = 28$ nm$^{-1}$\cite{hosokawa}.

The likely physics behind the de Gennes narrowing is straightforward.
The half-width of $S(k,\omega)$ is inversely proportional to the
lifetime of a density fluctuation of wave number $k$.  If that
$k$ coincides with the principal peak in the structure factor,
a density fluctuation will be in phase with the natural wavelength
of the liquid structure, and should decay slowly, in comparison
to density fluctuations at other wavelengths.  This is indeed the
behavior observed both in our simulations and in experiment.

In further support of this picture, we may attempt to
describe these fluctuations by a very oversimplified Langevin
model.  
We suppose that the Fourier component $\rho({\bf k}, t)$ 
[eq.\ (\ref{eq:rhokt})] is governed by a 
Langevin equation 
\begin{equation}
\dot{\rho}({\bf k},t) = -\xi\rho({\bf k}, t) + \eta(t).
\label{eq:langevin}
\end{equation}
Here the dot is a time derivative, $\xi$ is a constant, and
$\eta(t)$ is a random time-dependent ``force'' which has ensemble
average $\langle \eta\rangle = 0$ and correlation function
$\langle \eta(t)\eta^*(t^\prime)\rangle = A\delta(t - t^\prime)$.  
Eq.\ (\ref{eq:langevin}) can be solved by standard 
methods (see, e. g., Ref.\ \cite{hansen} for a related example), 
with the result (for sufficiently large t)
\begin{equation}
\langle \rho({\bf k}, t)\rho^*({\bf k}, t + \tau)\rangle =
\frac{\pi A}{\xi}\exp(-|\tau|\xi).
\label{eq:langevin1}
\end{equation}
According to eq.\ (\ref{eq:defskw}), $S(k,\omega)$ is, to within a
constant factor, the frequency Fourier transform of this expression,
i. e.
\begin{equation}
S(k, \omega) \propto \int_{-\infty}^\infty(\pi A/\xi)\exp(i\omega \tau)
\exp(-|\tau|\xi)d\tau, 
\end{equation}
or, on carrying out the integral,
\begin{equation}
S(k, \omega) \propto \frac{\pi A}{\omega^2 + \xi^2}.
\label{eq:langevin2}
\end{equation}
This is a Gaussian function centered at $\omega = 0$ and of
half-width $\xi$.   On the other hand, the static structure
factor 
\begin{equation}
S(k) \propto Lim_{\tau \rightarrow 0}
\langle\rho({\bf k}, t)\rho^*({\bf k}, t + \tau)\rangle 
= \frac{\pi A}{\xi}.
\end{equation}
Thus, if the constant $A$ is independent of ${\bf k}$, the
half-width $\xi$ of the function 
$S(k, \omega)$ at wave number ${\bf k}$ 
is inversely proportional to the static structure $S(k)$.  This
prediction is consistent with the ``de Gennes narrowing'' 
seen in our simulations and in experiment\cite{hosokawa}.

To summarize, there is overall a striking similarity
in the shapes of the experimental and calculated
curves for $S(k, \omega)/S(k)$ both in the hydrodynamic regime
and at larger values of $k$.

\subsubsection{$a$-Ge}

We have also calculated the dynamic structure for our sample
of $a$-Ge at $T = 300 K$.  The results
for the ratio $S(k, \omega)/S(k)$ are shown in Figs.\ 5 and 6
for a range of $k$ values, and, over a broader range of
$\omega$, in Fig.\ 7.  Once again, both
$S(k, \omega)$ and $S(k)$ are averaged over different values of
${\bf k}$ of the same length, as described above.  We have 
incorporated a resolution function
of width $\hbar\omega_0 = 2$ meV
into $S(k, \omega)$.   This  width is a rough estimate for the
experimental resolution function in the measurements
of Maley {\it et al}\cite{maley}; we 
assume it to be smaller than the liquid case because the 
measured width of the central peak in $S(k,\omega)/S(k)$
for $a$-Ge is quite small.

Ideally, our calculated $S(k, \omega)$ should be compared to
the measured one.  However, the published measured quantity
is not $S(k, \omega)$ but is, instead, based on 
a modified dynamical structure
factor, denoted $G(k, \omega)$, and related to $S(k, \omega)$
by\cite{maley}
\begin{equation}
G(k, \omega) = \left(\frac{C}{k^2}\right)
\left(\frac{\hbar\omega}{n(\omega, T)+1}\right)S(k, \omega).
\label{eq:modskw}
\end{equation}
Here $C$ is a k- and $\omega$-independent constant, and
$n(\omega, T) = 1/[e^{\hbar\omega/k_BT} - 1]$ is the phonon 
occupation number for 
phonons of energy $E = \hbar\omega$ at temperature $T$.  
The quantity plotted
by Maley {\it et al}\cite{maley} is an average of $G(k, \omega)$
over a range of k values from 40 to 70 nm$^{-1}$.  These workers 
assume that this average is proportional to 
the vibrational density of states $n_{vib}(\omega)$.
The measured $n_{vib}(\omega)$ as obtained
in this way\cite{maley} is shown in Fig.\ 8 for two different
amorphous structures, corresponding to two different methods
of preparation and having differing degrees of disorder.

In order to compare our calculated $S(k,\omega)$ to experiment,
we use eq.\ (\ref{eq:modskw}) to infer $G(k,\omega)$, then
average over a suitable range of k.  However, in using
eq.\ (\ref{eq:modskw}), we use the classical form of the
occupation factor, $n(\hbar\omega) + 1 \approx k_BT/\hbar\omega$,
This choice is justified because we have calculated $S(k, \omega)$
using classical equations of motion for the ions.  We thus obtain
for the calculated vibrational density of states
\begin{equation}
n_{vib}(\omega) \approx  
\left(\frac{C\hbar^2}{k_BT}\right)\left(\frac{\omega^2}{k^2}\right)
S(k, \omega).
\label{eq:nvibc}
\end{equation}
In Fig.\ 8 we show two such calculated plots of
$n_{vib}(\omega)$, as obtained by averaging
eq.\ (\ref{eq:nvibc}) over two separate
groups of $k$'s, as indicated in the caption\cite{note}.  For
comparison, we also show $n_{vib}(\omega)$ for $a$-Ge as calculated
in Ref.\ \cite{kresse} directly from the Fourier transform of
the velocity-velocity autocorrelation function.  

The calculated plots for $n_{vib}(\omega)$ 
in Fig.\ 8 have some distinct structure, which arises
from some corresponding high frequency structure in
$S(k, \omega)$.   The plot of $n_{vib}(\omega)$ for the group
of smaller k's has two distinct peaks, near $8$ meV and $29$ meV,
separated by a broad dip with a minimum near 18 meV.
The plot corresponding to the group of larger k's has similar
structure and width, but the dip is less pronounced.  
The two experimental plots 
also have two peaks separated by a clear dip.  
The two maxima are found around $10$ and $35$ meV, while
the principal dip occurs near $16$ meV.  In addition, the overall
width of the two densities of states is quite similar.  

The reasonable agreement between the calculated
and measured $n_{vib}(\omega)$ suggests that
our {\em ab initio} calculation of $S(k,\omega)$ for
$a$-Ge is reasonably accurate.  The noticeable differences
probably arise from several factors.  First, there are several
approximations involved in going from the calculated and 
measured $S(k, \omega)$'s to the corresponding $n_{vib}(\omega)$'s,
and these may be responsible for some of the discrepancies.
Secondly, there may actually be differences between the particular
amorphous structures studied in the experiments, and the
quenched, then relaxed structure considered in the present
calculations.  (However, the similarities in the {\em static}
structure factors suggest that these differences are not
vast.)  Finally, our calculations are carried out over relatively
short times, using relatively few atoms; thus, finite-size and
finite-time effects are likely to produce some additional 
errors.  Considering all these factors, agreement between calculation
and experiment is quite reasonable.

Previous {\em ab initio} calculations for $a$-Ge\cite{kresse}
have also obtained a vibrational density of states, but this
is computed directly from the ionic velocity-velocity
autocorrelation function rather
than from the procedure described here.  The calculations
in Ref.\ \cite{kresse} 
do not require computing $S(k, \omega)$.  In the
present work, by contrast, we start from $S(k,\omega)$ (which
is calculated here for the first time in an
{\em ab initio} calculation for $a$-Ge), and we
work backwards to get $n_{vib}(\omega)$.  In principle, our
$S(k, \omega)$ includes all anharmonic effects on the vibrational
spectrum of $a$-Ge, though in extracting $n_{vib}(\omega)$ we
assume that the lattice vibrates harmonically about the metastable
atomic positions.  In Fig.\ 8, we also show the results of
Ref.\ \cite{kresse} for $n_{vib}(\omega)$ as obtained from this
correlation function.  They are quite similar to those obtained
in the present work, but have a somewhat deeper minimum between
the two principal peaks.

The quantity $n_{vib}(\omega)$ could, of course, also be calculated
directly from the force constant matrix, obtained by assuming
that the quenched configuration is a local energy minimum and
calculating the potential energy for small positional deviations
from that minimum using {\em ab initio} molecular dynamics.
This procedure has been followed for $a$-GeSe$_2$, for example,
by Cappelletti {\it et al}\cite{cappelletti}.  These
workers have then obtained $S(q, \omega)$ versus $q$ from
their $n_{vib}(\omega)$ at selected values of
$\omega$, within a one-phonon approximation.  However, as noted
above, the present work produces the full $S(k, \omega)$ and
thus has, in principle, more information than $n_{vib}(\omega)$.

\section{DISCUSSION AND CONCLUSIONS}

The present results show that {\em ab initio} molecular
dynamics can be used to calculate the dynamic structure
factor $S(k, \omega)$ for both liquid and amorphous semiconductors.
Although the accuracy of the calculated $S(k, \omega)$ is lower
than that attained for static quantities, such as $S(k)$, 
nonetheless it is sufficient for comparison to most experimental
features.  This is true even though our calculations are limited
to 64-atom samples and fewer than 20 ps of elapsed real time.

We have presented evidence that the calculated
$S(k, \omega)/S(k)$ in $\ell$-Ge agrees qualitatively with
measured by inelastic X-ray scattering\cite{hosokawa}, and that
the one calculated for $a$-Ge leads to a vibrational density of
states qualitatively similar to the quoted 
experimental one\cite{maley}.
Since such calculations are thus shown to be feasible,
our work should spur further numerical studies, with longer
runs on larger samples, to obtain even more detailed information.
Furthermore, we can use these dynamical simulations to probe
the underlying processes at the atomic scale which give rise
to specific features in the measured and calculated $S(k,\omega)$.

\section{ACKNOWLEDGEMENTS}

This work has been supported by NASA, Division of Microgravity
Sciences, through grant NCC8-152, and by
NSF grants DMR01-04987 (DS) and CHE 01-11104 (JDC).
Calculations were
carried out using the Beowulf Cluster at 
the Ohio Supercomputer Center, with the help of a grant of time.
We are very grateful to Prof. David H. Matthiesen for his continual 
support and encouragement through the course of this work,
and to Sergey Barabash for many valuable conversations. Jeng-Da Chai 
wishes to thank Lan Bi for her endless support.

\newpage

\begin{center}

{\bf Figure Captions}

\end{center}

\begin{enumerate}

\item  Static structure factor S(k) for
$\ell$-Ge at $T = 1250 K$, just above the experimental
melting temperature.  Full curve: present work, as calculated
using the generalized gradient approximation (GGA; see text).
Dashed curve: present work, but using the local density
approximation (LDA; see text).  Open circles: measured S(k) near 
$T = 1250$ K, as given in Ref.\ \cite{waseda}. 

\item Full curve: Calculated S(k) for $a$-Ge at $T = 300 K$,
as obtained using the GGA.  Structure is prepared as described in 
the text.  Open circles: measured S(k) for $a$-Ge at $T = 300 K$, 
as given in Ref.\ \cite{etherington}.

\item Calculated ratio of dynamic structure factor
$S(k, \omega)$ to static structure factor $S(k)$ for $\ell$-Ge
at $T = 1250 K$ for several values of $k$, plotted as a function
of $\omega$, calculated using {\em ab initio} molecular
dynamics with a MD time step of 10 fs.
For clarity, each curve has been vertically displaced by 0.05 units 
from the curve below.
In each case, the plotted curve is obtained 
by averaging both the calculated
$S({\bf k}, \omega)$ and the calculated $S({\bf k}) $
over all values of ${\bf k}$ of the same
length.  We also incorporate a Gaussian resolution function of
half-width $\hbar\omega_0 = 2.5$ meV, as in eqs.\ (\ref{eq:resol})
and (\ref{eq:resol1}).  This value of $\omega_0$ is
chosen to equal the
quoted experimental resolution\cite{hosokawa}.

\item Same as in Fig.\ 3, but without the resolution function.

\item Calculated $S(k,\omega)/S(k)$ for $a$-Ge at $T = 300$ K
at $k \leq 35$ nn$^{-1}$, plotted as a function of $\omega$
Again, each curve has been vertically displaced by appropriate
amounts from the one below it, as evident from the Figure, 
and both $S(k,\omega)$ and $S(k)$ have
been plotted after an average over all ${\bf k}$'s of the
same length.   
We also incorporate a Gaussian resolution
function of half-width $\hbar\omega_0 =2$ meV.  This value
is chosen to give the best results for $n_{vib}(\omega)$ as
measured by  Ref.\ \cite{maley}.  The time step here is
10 fs.

\item Same as Fig.\ 5, but at $k \geq 35$ nm$^{-1}$.

\item Calculated $S(k,\omega)/S(k)$ as in Figs.\
5 and 6 but including higher frequencies $\omega$.
The vertical displacements are the same as in Figs.\ 5 and 6.
At $\hbar\omega = 60$ meV, the curves are arranged vertically
in order of increasing frequency.

\item Full curve: calculated vibrational density of states
$n_{vib}(\omega)$, in units of $10^{-3}$ states/meV.  
$n_{vib}(\omega)$ is obtained from the resolution-broadened 
$S(k,\omega)$ and $S(k)$ of the previous Figure, using
the formula 
$n_{vib}(\omega) = \langle G_{calc}(k, \omega)\rangle$,
where $G_{calc}(k, \omega)$ is given by eq.\ (16), and
the averaging is carried out over
the three magnitudes of $k$ near $40$ nm$^{-1}$
for which we have computed $S(k, \omega)$.
Dashed curve: same as full curve, but calculated
by averaging over the
six magnitudes of $k$ near $90$ nm$^{-1}$
for which we have computed $S(k, \omega)$.
The open circles and open stars denote
the measured $n_{vib}(\omega)$, as reported in Ref.\ \cite{maley}
for two forms of $a$-Ge.  Finally, the open diamonds denote
$n_{vib}(\omega)$ as calculated in Ref.\ \cite{kresse} from the
ionic velocity-velocity autocorrelation function (dot-dashed 
curves). 
In all plots except that of Ref.\ \cite{kresse}, 
$n_{vib}(\omega)$ is normalized 
so that $\int_0^{\omega_{max}}n(\omega)d(\hbar\omega) = 1$.
$\omega_{max}$ is the frequency at which
$n_{vib}(\omega) \rightarrow 0$, and is estimated from
this Figure by extrapolating the right hand parts of the solid
and dashed curves linearly to zero.  

\end{enumerate}

\newpage
\begin{figure}
\includegraphics{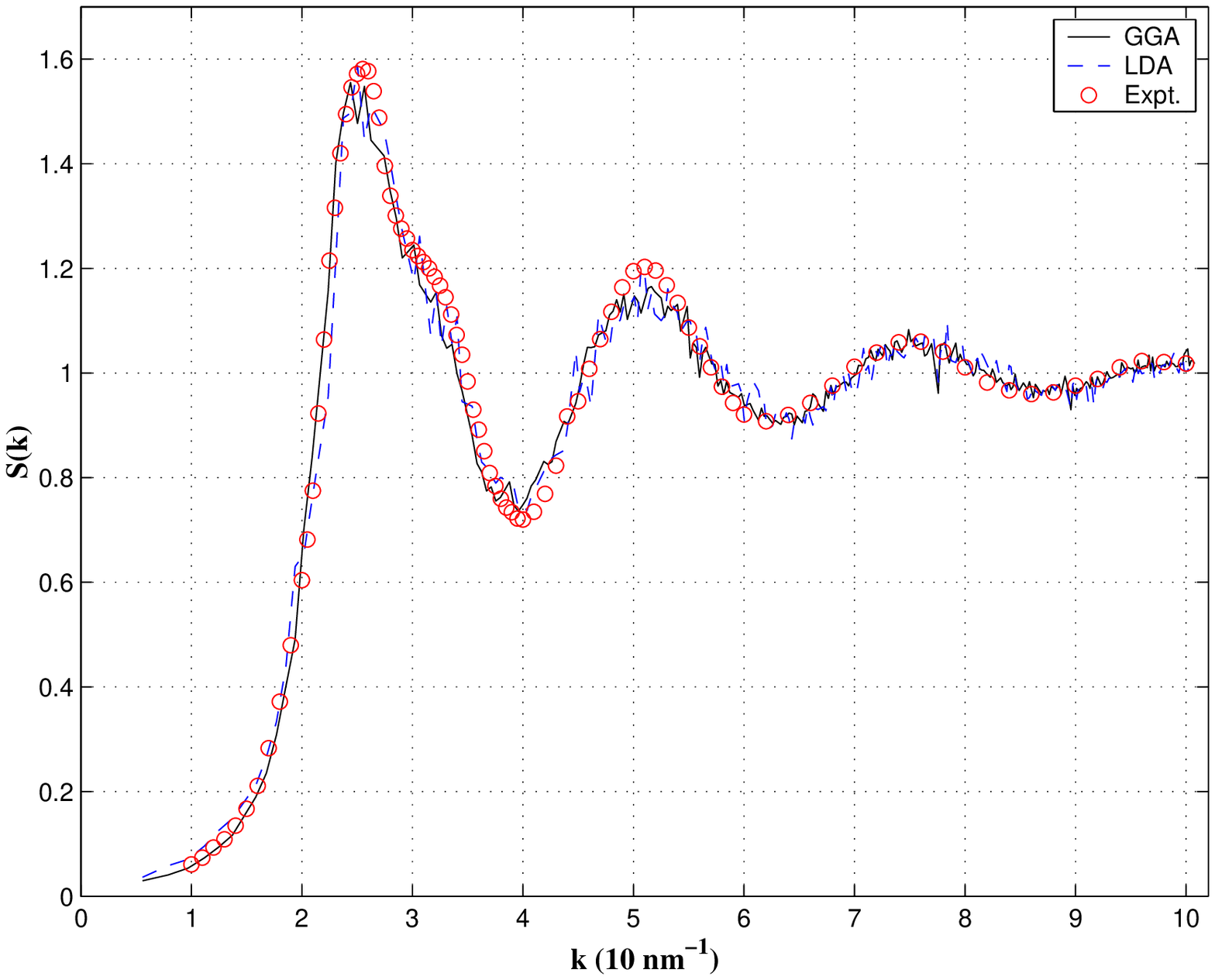}%
\caption{\label{1}}
\end{figure}
\begin{figure}
\includegraphics{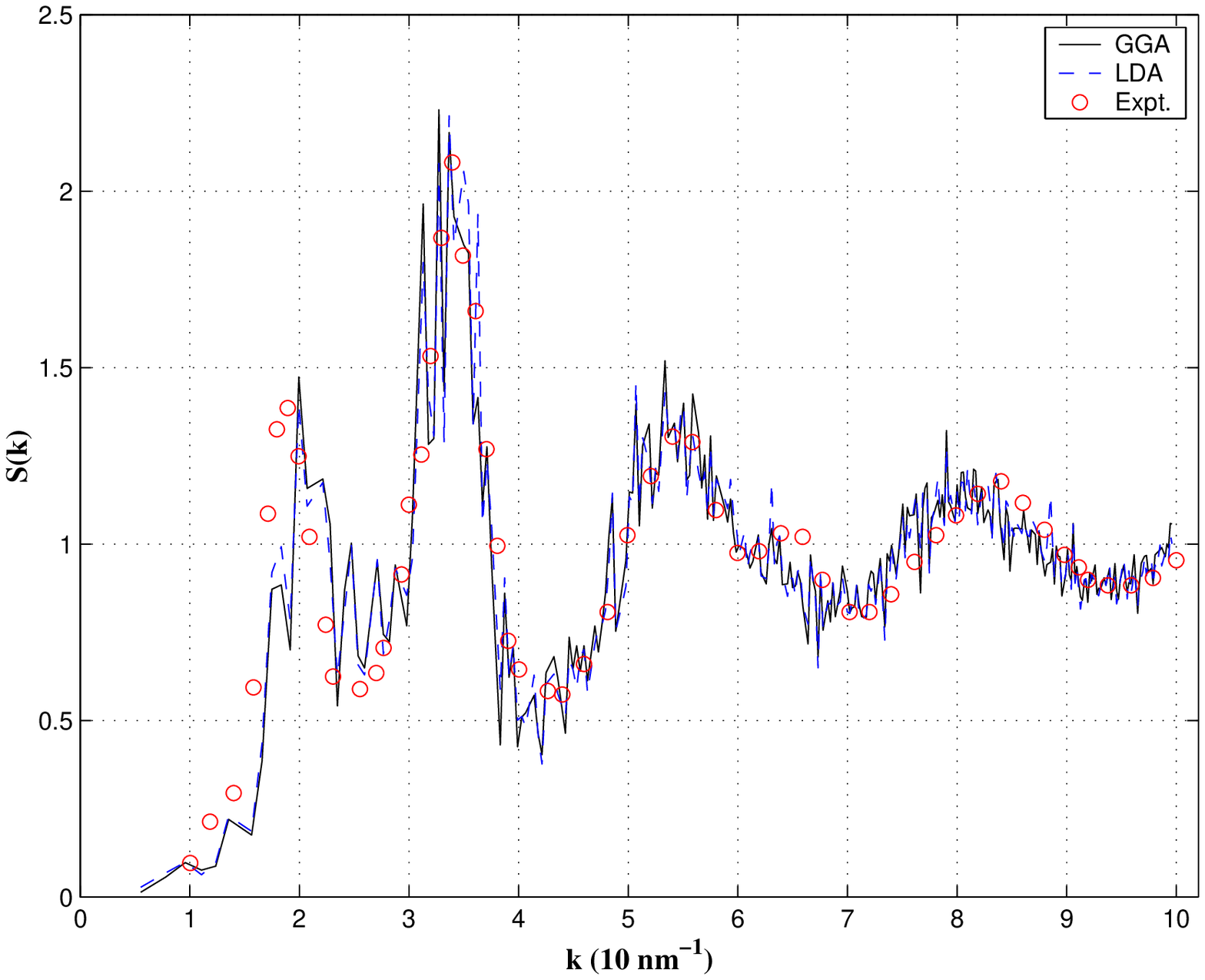}%
\caption{\label{2}}
\end{figure}
\begin{figure}
\includegraphics{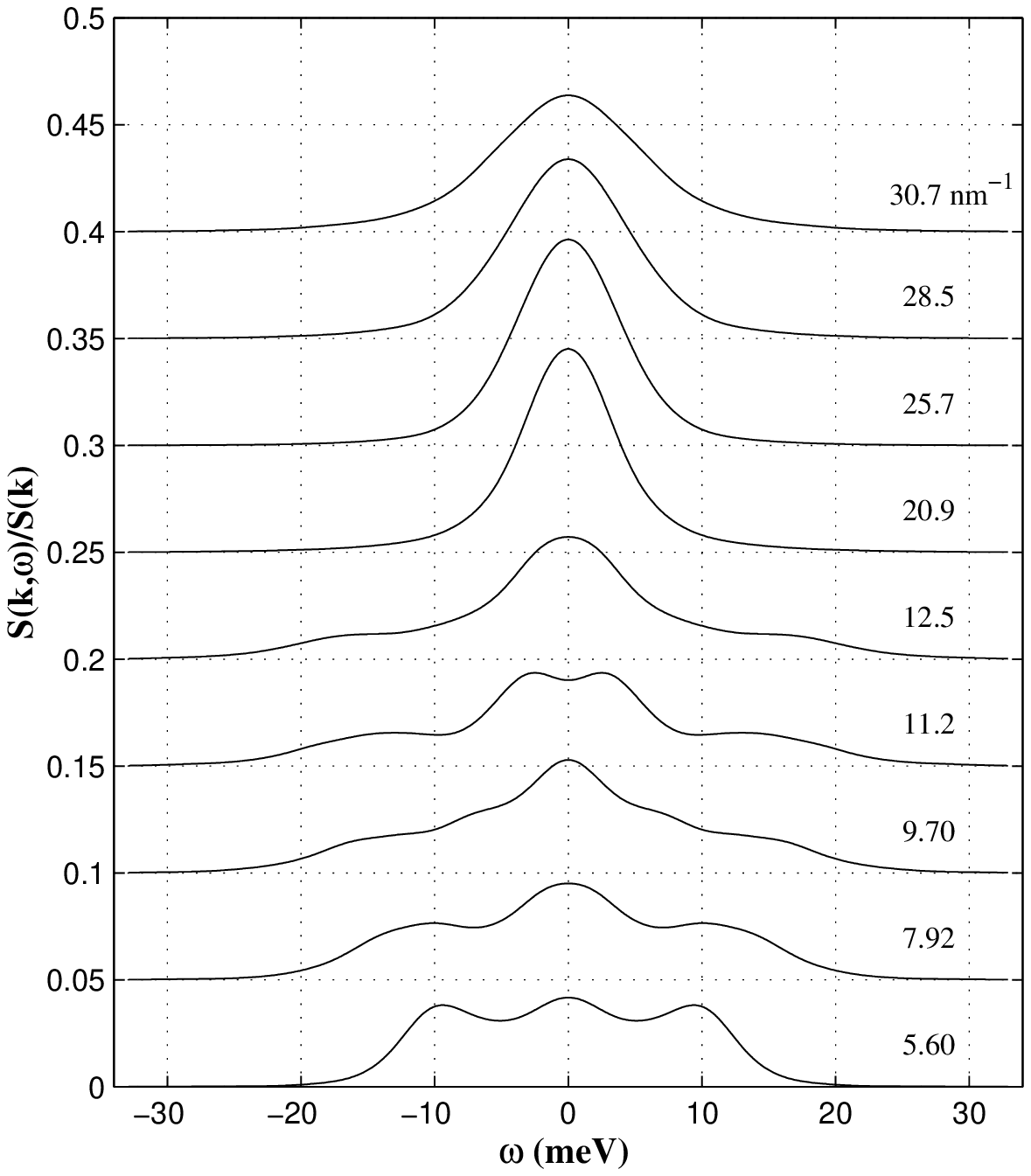}%
\caption{\label{3}}
\end{figure}
\begin{figure}
\includegraphics{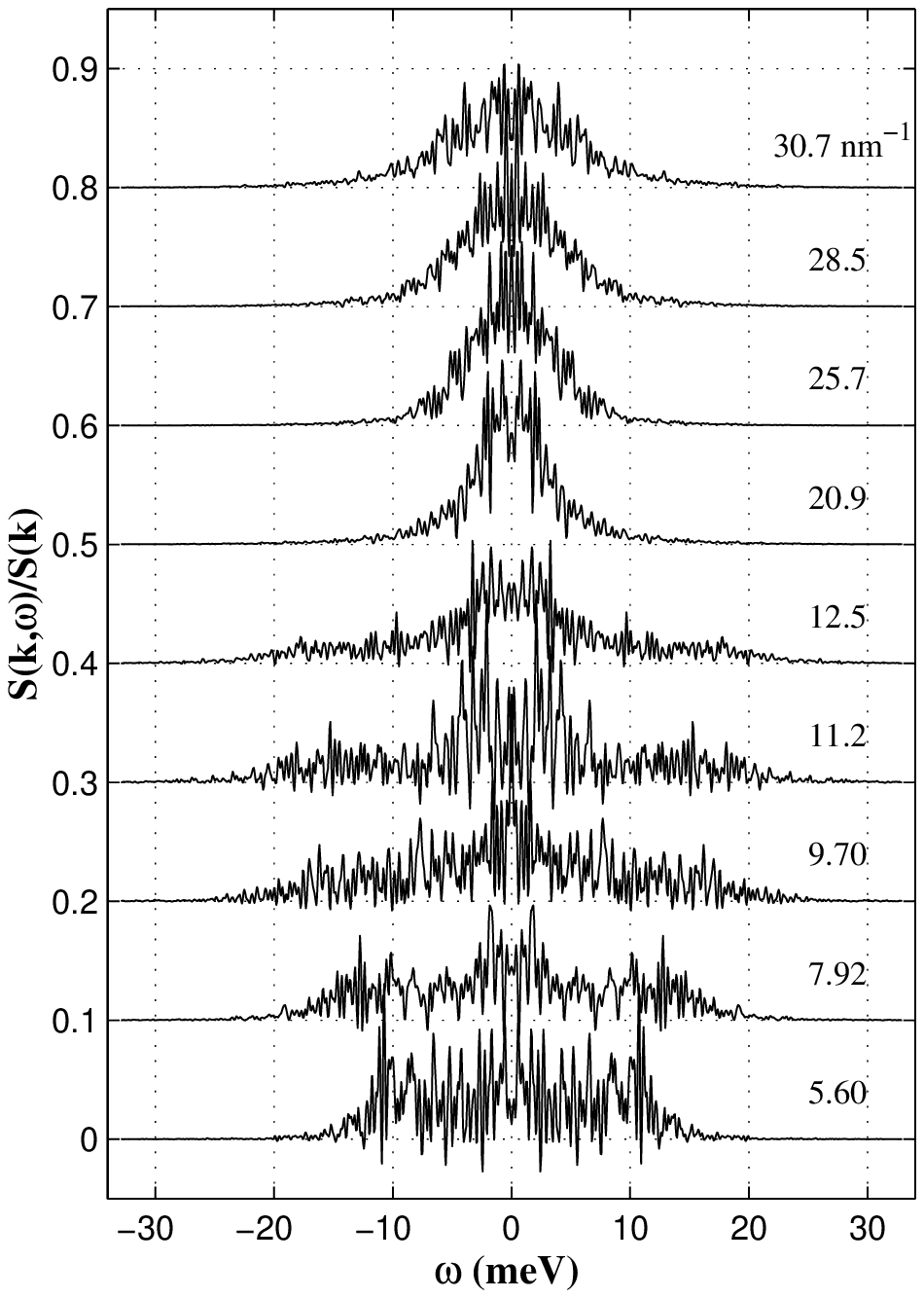}%
\caption{\label{4}}
\end{figure}
\begin{figure}
\includegraphics{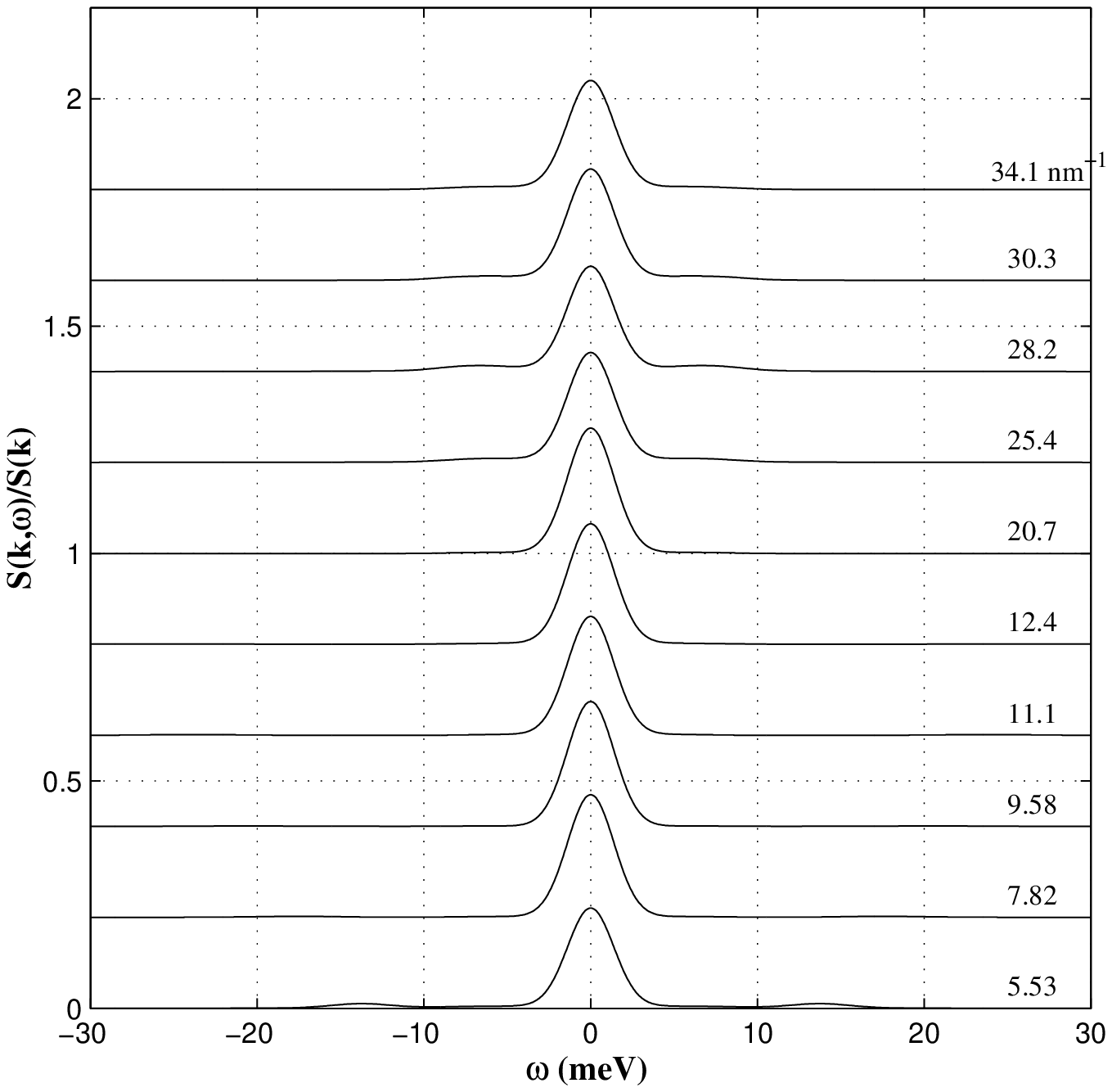}%
\caption{\label{5}}
\end{figure}
\begin{figure}
\includegraphics{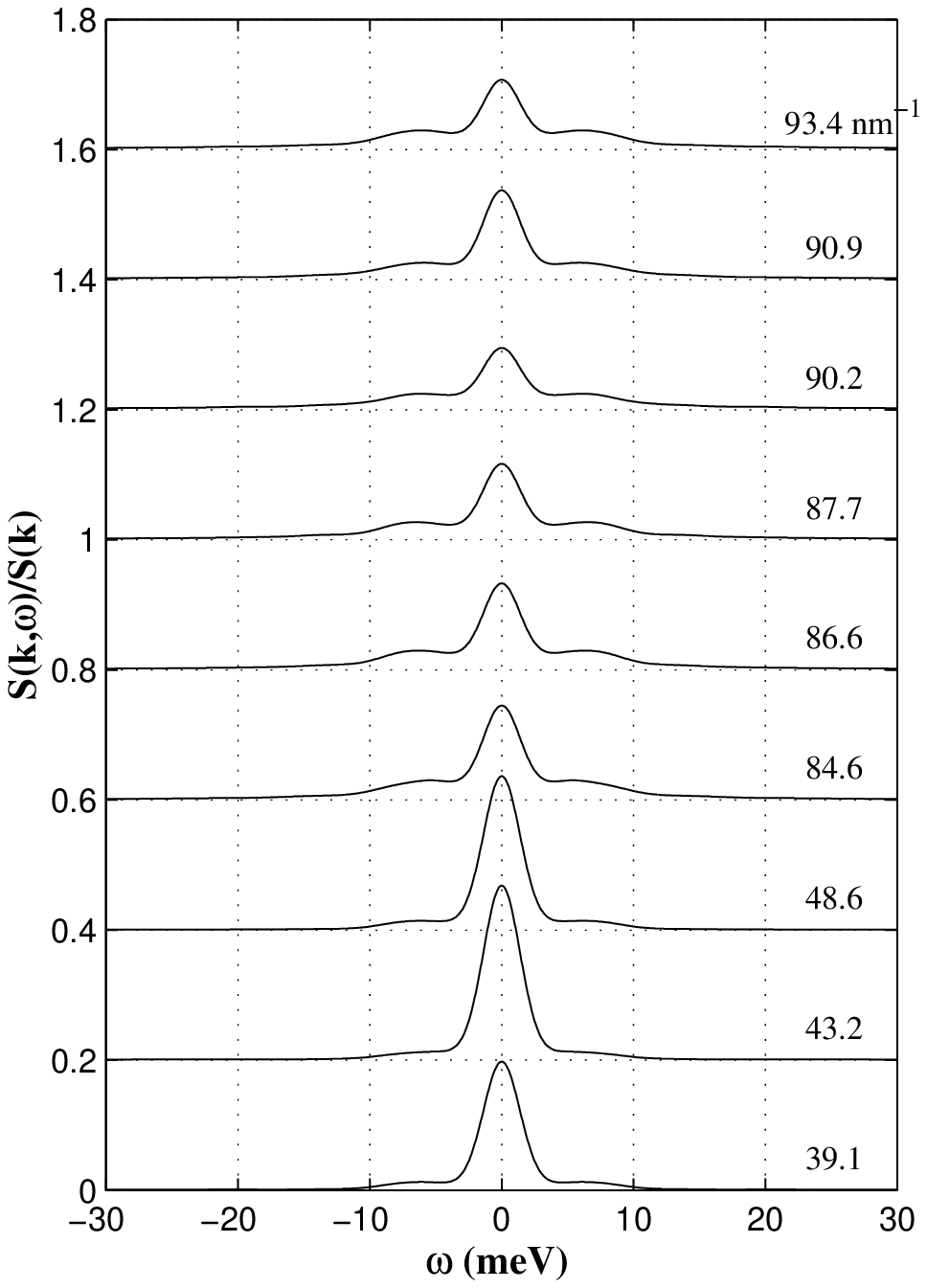}%
\caption{\label{6}}
\end{figure}
\begin{figure}
\includegraphics{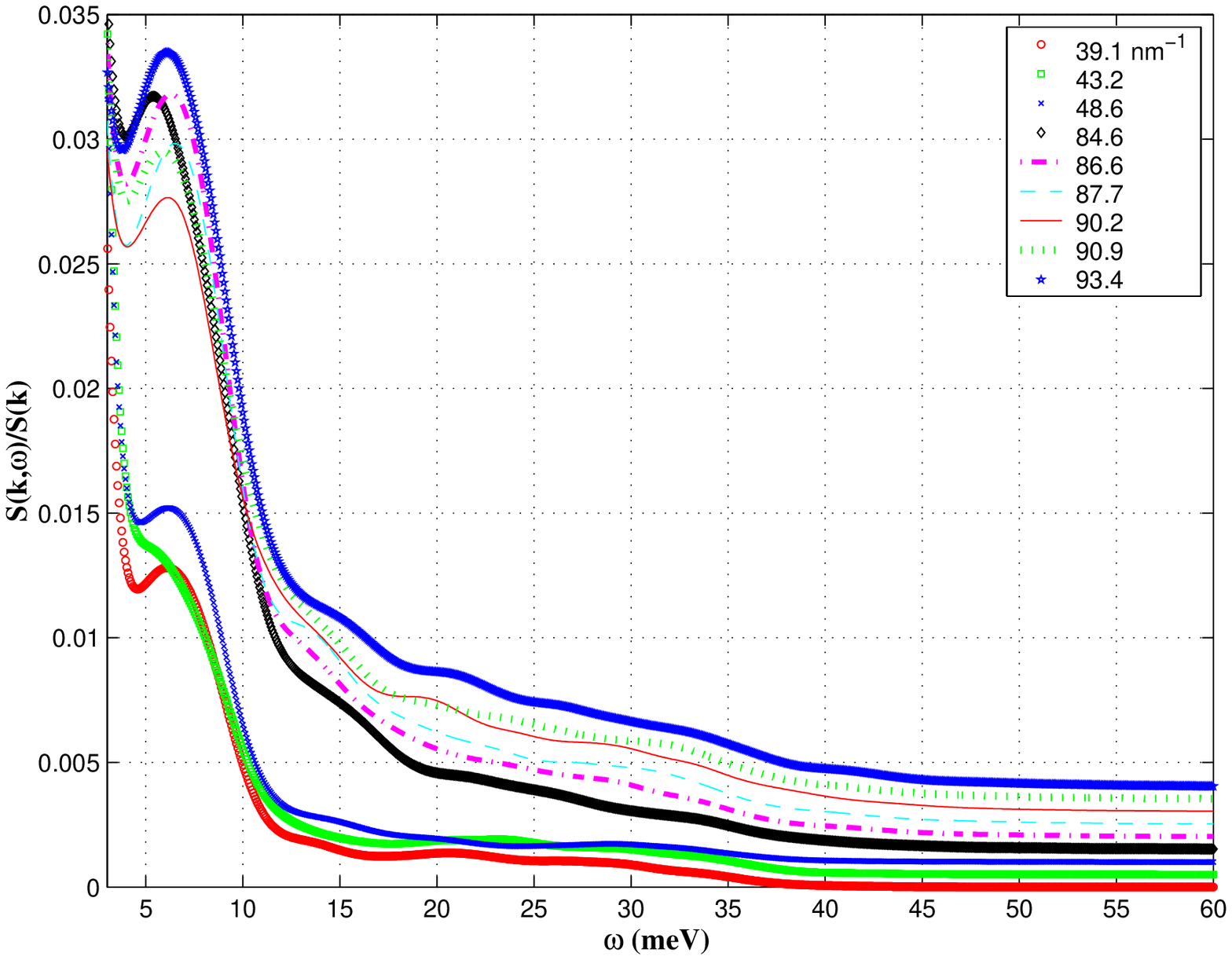}%
\caption{\label{7}}
\end{figure}
\begin{figure}
\includegraphics{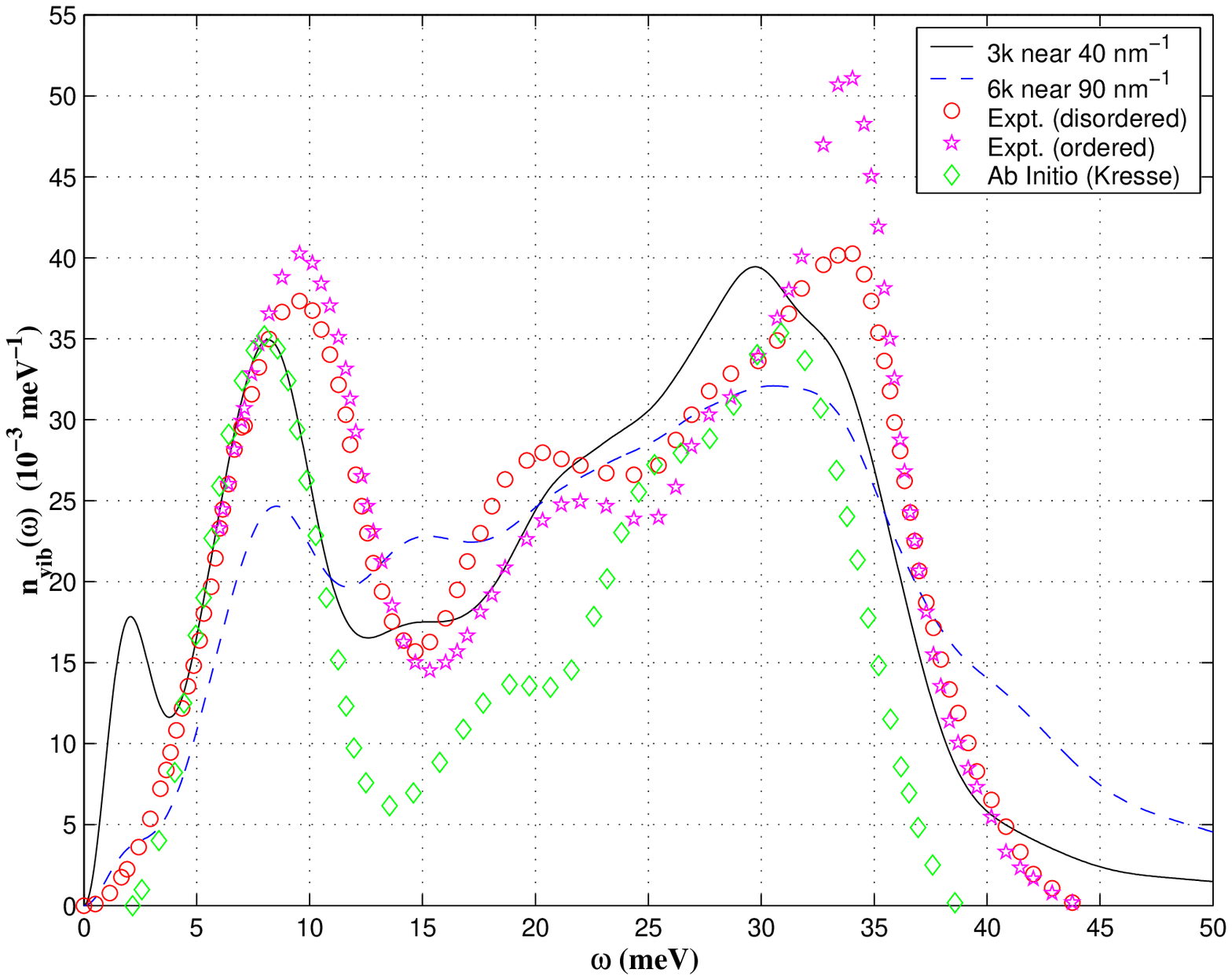}%
\caption{\label{8}}
\end{figure}


\begin{thebibliography}{99}
\bibitem{compare} Our method is very similar to that used
in Ref.\ \cite{kresse} to treat $\ell$-Ge and $a$-Ge, but
with the following differences: (i) we use both
the GGA and the LDA, while
Ref.\ \cite{kresse} uses only the LDA; 
(ii) we include slightly more bands (161 as
compared to 138); and (iii) most of our calculations are
carried out using a 10 fs time step, rather than the 3 fs
step used in Ref.\ \cite{kresse}.  Our $a$-Ge structure may
also differ slightly from that of Ref.\ \cite{kresse}, since
we use a somewhat faster quench rate ($\sim 3 \times 10^{14}$ K/sec
rather than $1.7 \times 10^{14}$ K/sec).  Both codes use
the conjugate gradient method to find energy minima and
the same Vanderbilt ultrasoft pseudopotentials.
\end{thebibliography}
\end{document}